\documentclass[%
 aip,
 jmp,%
 amsmath,amssymb,
 reprint,%
]{revtex4-1}

\usepackage{graphicx}
\usepackage{dcolumn}
\usepackage{bm}

\begin{document}

\preprint{AIP/123-QED}

\title{Influence of trap location on the efficiency of trapping in dendrimers and regular hyperbranched polymers}

\author{Yuan Lin}
\author{Zhongzhi Zhang}
\email{zhangzz@fudan.edu.cn}
\homepage{http://www.researcherid.com/rid/G-5522-2011}

\affiliation {School of Computer Science, Fudan University,
Shanghai 200433, China}

\affiliation {Shanghai Key Lab of Intelligent Information
Processing, Fudan University, Shanghai 200433, China}

\date{\today}

\begin{abstract}
The trapping process in polymer systems constitutes a fundamental mechanism for various other dynamical processes taking place in these systems. In this paper, we study the trapping problem in two representative polymer networks, Cayley trees and Vicsek fractals, which separately model dendrimers and regular hyperbranched polymers. Our goal is to explore the impact of trap location on the efficiency of trapping in these two important polymer systems, with the efficiency being measured by the average trapping time (ATT) that is the average of source-to-trap mean first-passage time over every staring point in the whole networks. For Cayley trees, we derive an exact analytic formula for the ATT to an arbitrary trap node, based on which we further obtain the explicit expression of ATT for the case that the trap is uniformly distributed. For Vicsek fractals, we provide the closed-form solution for ATT to a peripheral node farthest from the central node, as well as the numerical solutions for the case when the trap is placed on other nodes. Moreover, we derive the exact formula for the ATT corresponding to the trapping problem when the trap has an uniform distribution over all nodes. Our results show that the influence of trap location on the trapping efficiency is completely different for the two polymer networks. In Cayley trees, the leading scaling of ATT increases with the shortest distance between the trap and the central node, implying that trap's position has an essential impact on the trapping efficiency; while in Vicsek fractals, the effect of location of the trap is negligible, since the dominant behavior of ATT is identical, respective of the location where the trap is placed. We also present that for all cases of trapping problems being studied, the trapping process is more efficient in Cayley trees than in Vicsek fractals. We demonstrate that all differences related to trapping in the two polymer systems are rooted in their underlying topological structures.
\end{abstract}


\maketitle


\section{Introduction}

As a paradigmatic dynamical process, trapping describes and characterizes various other important physical processes on complex systems, e.g., page search or access in the World Wide Web~\cite{HwLeKa12,HwLeKa12E,BeCoMoSuVo05,BeLoMoVo11}. The trapping problem constitutes an integral primary problem of random walks, defined as a kind of isotropic random walks with a perfect trap located at a given position, absorbing all particles that visit it~\cite{Mo69}. The highly desirable quantity related to the trapping problem is the trapping time (TT), also known as mean first-passage time (MFPT)~\cite{Re01,NoRi04,BeCoMo05,CoBeKl07,CoBeTeVoKl07}, which represents the expected time for a walker starting off from a source node to arrive at the trap for the first time. The average of trapping time over all starting nodes is defined as the average trapping time (ATT), which is very important and useful in related fields, since it is often used as a quantitative indicator measuring the efficiency of trapping process.

In view of the theoretical and impractical relevance, determining ATT in diverse systems has received a tremendous amount of attention within the scientific community. Thus far, trapping problem has been extensively studied for a lot of complex systems, such as regular lattices with different dimensions~\cite{Mo69,BaKl98JPC,GLKo05,GLLiYoEvKo06,CaAb08}, the treelike $T-$ fractals~\cite{KaRe89,Ag08,HaRo08,ZhLiZhWuGu09,LiWuZh10,ZhWuCh11}, the small-world uniform recursive trees~\cite{ZhQiZhGoGu10,ZhLiLiCh11}, the Sierpinski gaskets~\cite{KaBa02PRE,KaBa02IJBC,BeTuKo10}, as well as fractal~\cite{ZhXiZhGaGu09,ZhXiZhLiGu09,ZhYaGa11,TeBeVo09} or non-fractal~\cite{KiCaHaAr08,ZhQiZhXiGu09,ZhZhXiChLiGu09,AgBu09,AgBuMa10,MeAgBeVo12,YaZh13} even modular~\cite{ZhLiGoZhGuLi09,ZhYaLi12} scale-free graphs. These works have unveiled some nontrivial influences of certain particular structural properties on the leading behaviors of ATT for trapping problem performed in these different systems.

In addition to aforementioned systems, trapping in polymer systems~\cite{GuBl05} is also a fundamental topic due to its wide range of applications, including lighting harvesting in antenna systems~\cite{BaKlKo97,BaKl98,BeHoKo03,BeKo06,Ag11}, energy or exciton transport in polymer systems~\cite{SoMaBl97,BlZu81}, and so on. Among various polymer systems, Cayley trees~\cite{CaCh97,ChCa99} and Vicsek fractals~\cite{Vi83,BlJuKoFe03,BlFeJuKo04} are two important classes modeling separately dendrimers and regular hyperbranched macromolecules, both of which have been and continue to be active subjects of research in numerous fields~\cite{SuHaFrMu99,SuHaFr00,GaFeRa01,BiKaBl01,HeMaKn04,MuBiBl06,MuBl11,Vi84,JaWu92,JaWu94,StFeBl05,ZhZhChYiGu08,Vo09,ZhWjZhZhGuWa10,JuvoBe11}. Recently trapping problem in Cayley trees~\cite{BeHoKo03,BeKo06,WuLiZhCh12} and Vicsek fractals~\cite{WuLiZhCh12} with a deep trap at the central node has been studied, uncovering how the underlying structures of the two polymer networks affect the efficiency of this special trapping process on them. However, previous works~\cite{TeBeVo09,CoBeMo07,TeBeVo11,LiJuZh12} have shown that for trapping in a network (e.g., scale-free network~\cite{TeBeVo09,LiJuZh12}), the trapping efficiency may rely on the position of the trap. Then, two interesting open questions arise naturally for Cayley trees and Vicsek fractals: What is the impact of trap's location on the behavior for ATT in these two polymer networks? And how the ATT scales with the system size when the trap is uniformly distributed over the networks?

In this paper, in order to explore the role of trap's position on the trapping efficiency in polymer systems, we study the trapping problem on Cayley trees~\cite{CaCh97,ChCa99} and Vicsek fractals~\cite{Vi83,BlJuKoFe03,BlFeJuKo04}, both of which grow in an iterative manner and have a treelike structure that allow for analytically treating most cases of trapping processes occurring on them, obtaining explicit expressions for the ATT characterizing the trapping processes. We focus on attacking two cases of trapping problems on the two polymer networks. In the first case, the trap is fixed on a particular node, while in the other case, the trap is distributed uniformly over all nodes in separate networks under consideration. We obtain analytical closed-form solutions or numerical solutions to the ATT for both cases.

Concretely, for Cayley trees, we derive an exact formula for ATT when the trap is located at an arbitrary node, the leading behavior of which depends on the shortest distance between the trap and the central node: the longer the distance, the higher the ATT, indicating that the position of trap has a substantial effect on the trapping efficiency. While for Vicsek fractals, we obtain an exact analytical solution to ATT when the trap is a peripheral node, as well the numerical solutions to ATT when the trap is placed on another node, with both solutions having the same dominant scaling, implying that the influence of trap's location on the ATT is negligible. For the second trapping problem, by using different approaches, we deduce accurate expressions for the ATT for both networks, which show the trapping efficiency of Cayley trees is much higher than that of Vicsek fractals. We present that their particular structures are responsible for the difference of results obtained for the two networks. This work deepens the understanding of trapping problems in the two representative polymer networks.

\section{Network constructions and properties}

In this section, we introduce the two representative polymer networks---Cayley trees and Vicsek fractals---and their structural properties. Both networks are constructed in a deterministically iterative manner. 

\subsection{Cayley trees}

\begin{figure}
\begin{center}
\includegraphics[width=0.8\linewidth,trim=0 0 0 0]{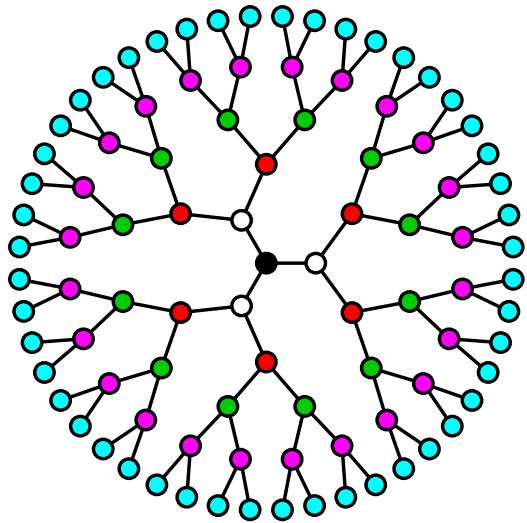}
\caption{(Color online) Structure of the a particular Cayley tree $C_{3,5}$. }  \label{Cayley}
\end{center}
\end{figure}

Cayley trees~\cite{CaCh97,ChCa99} after $g$ iterations, denoted by $C_{m,g}$ $(m\geq 3,  g\geq 0)$, are built in the following iterative way. For $g=0$, $C_{m,0}$ comprises only a central node. For $g=1$, $m$ nodes are generated connecting the central node to form $C_{m,1}$, with the $m$ single-degree nodes constituting the peripheral nodes of $C_{m,1}$. For any $g > 1$, $C_{m,g}$ is obtained from $C_{m,g-1}$: for each peripheral node of $C_{m,g-1}$, we add $m-1$ new nodes and link them to the peripheral node. All the new introduced nodes at this stage become the peripheral nodes of $C_{m,g}$. Figure~\ref{Cayley} illustrates the construction process for a specific Cayley tree $C_{3,5}$.

By construction, one can easily check that at each iteration $i$ ($i \geq 1$) the number of newly created nodes is $N_{i}(g)=m(m-1)^{i-1}$. Thus, for $C_{m,g}$, the number of peripheral nodes and the total number of nodes are
\begin{eqnarray}\label{PeriC}
\bar N_{g}=N_g(g)=m(m-1)^{g-1}\,
\end{eqnarray}
and
\begin{equation}\label{NodeC}
N_g=1 + \sum_{i=1}^{g}N_{i}(g)=\frac{m(m-1)^g-2}{m-2}\,,
\end{equation}
respectively.
Then, the total number of edges in $C_{m,g}$ is
\begin{equation}\label{EegeC}
E_g=N_g-1=\frac{m(m-1)^g-m }{m-2}\,.
\end{equation}

It should mentioned that although Cayley trees exhibit an obvious self-similar structure, they are nonfractal since they have an infinite fractal dimension.

\subsection{Vicsek fractals}

\begin{figure}
\begin{center}
\includegraphics[width=0.85\linewidth,trim=0 0 0 0]{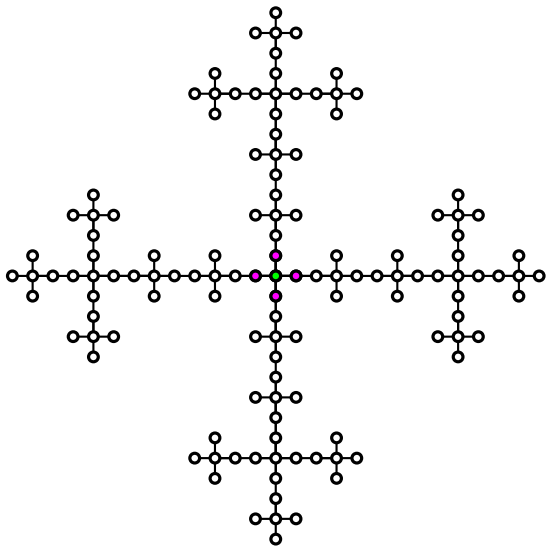}
\caption{Illustration of the first several iterative processes of a
particular Vicsek fractal, $V_{4,3}$. } \label{Vicsek}
\end{center}
\end{figure}

Vicsek fractals~\cite{Vi83,BlJuKoFe03,BlFeJuKo04} after $g$ iterations, denoted by $V_{f,g}$ ($f\geq3, g\geq 0$), are constructed in a different iterative way from that of Cayley trees. For $g=0$, $V_{f,0}$ consists of an isolated node without any edge. For $g=1$, $f$ new nodes are generated with each being connected to the node in $V_{f,0}$ to form $V_{f,1}$, which is exactly a star. For $g\geq2$, $V_{f,g}$ is obtained from $V_{f,g-1}$. To obtain $V_{f,g}$, we introduce $f$ new identical copies of $V_{f,g-1}$ and arrange them around the periphery of the original $V_{f,g-1}$. Then we add $f$ new edges, each of them connecting a peripheral node in one of the $f$ corner copy structures and a peripheral node of the original central structure, where a peripheral node is a node farthest from the central node. Figure~\ref{Vicsek} shows the structure of a special Vicsek fractal $V_{4,3}$.
By construction, at each generation the number of the nodes increases by a factor of $f+1$; therefore, the total number of nodes of $V_{f,g}$ is $N_g=(f+1)^g$, and the total number of edges in $V_{f,g}$ is $E_g=N_g-1=(f+1)^g-1$. In contrast to the Cayley trees, Vicsek fractals are fractal objects with the fractal dimension being equal to $\log_{3}(f+1)$.

We continue to study some properties of the Vicsek fractals, which are useful for the following text. Let $L_g$ denote the diameter of $V_{f,g}$. It is easy to see that $L_g$ satisfies recursive relation $L_g=3L_{g-1}+2$, which together with the initial condition $L_1=2$ leads to $L_g=3^g-1$. Actually, for two nodes in $V_{f,g}$, if their shortest distance is equal to the diameter $L_g$, then the shortest path connecting the two nodes must include the central node of $V_{f,g}$, and these two nodes are two peripheral nodes of $V_{f,g}$.
Let $\bar N_g$ denote the number of peripheral nodes of $V_{f,g}$. Evidently, $\bar N_g$ satisfies recursive relation $\bar N_g=(f-1)\bar N_{g-1}$. Considering $\bar N_1=f$, we have
\begin{eqnarray}\label{X01}
\bar N_g=f(f-1)^{g-1}\,.
\end{eqnarray}

\begin{figure}
\begin{center}
\includegraphics[width=0.65\linewidth,trim=0 0 0 0]{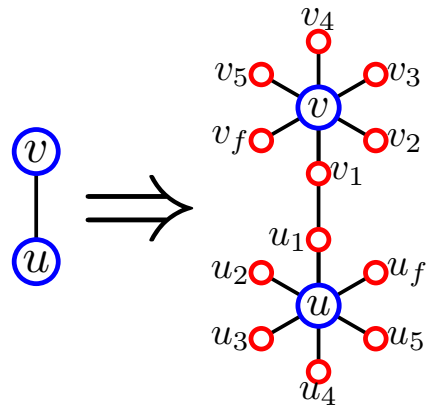}
\end{center}
\caption[kurzform]{(Color online) Another construction method
of Vicske fractals. $u$ and $v$ are two adjacent nodes in  $V_{f,g-1}$. At iteration $g$, each of them gives rise to $f$ new neighbors, denoted by $u_1$, $u_2$,$\ldots$, $u_f$, and $v_1$, $v_2$,$\ldots$, $v_f$, respectively. Since $u$ and $v$ are directly connected by an edge in  $V_{f,g-1}$, two of their new neighboring nodes (e.g.,
$u_1$ and $v_1$) are connected to each other by a new edge. }\label{Const2}
\end{figure}

In addition to the above replication and connection operations, Vicsek fractals can also be alternatively constructed using another method~\cite{BlFeJuKo04} as illustrated in  Fig.~\ref{Const2}. Suppose one has $V_{f, g-1}$. Then, $V_{f, g}$ can be obtained from $V_{f, g-1}$ by performing the following steps. First, for each node in $V_{f, g-1}$, $f$ new nodes are generated and linked to the old node. Then, for each pair of adjacent nodes, $u$ and $v$ in $V_{f, g-1}$, a new edge is added between two of their new neighboring nodes. Moreover, each new node generated at generation $g$ can have at most one new neighbor born at the same generation.

According to the second construction way, we can categorize the nodes in $V_{f,g}$. Let $\Lambda_g$ represent the set of nodes of $V_{f,g}$, and ${\bar\Lambda}_g$ the set of those nodes of $V_{f,g}$, which are created at iteration $g$. Obviously, $\Lambda_g=\Lambda_{g-1}+\bar\Lambda_g$. Moreover, we can classify the set $\bar\Lambda_g$ into two subsets $\bar\Lambda_g^{(1)}$ and $\bar\Lambda_g^{(2)}$, such that $\bar\Lambda_g=\bar\Lambda_g^{(1)}\cup\bar\Lambda_g^{(2)}$, where $\bar\Lambda_g^{(1)}$ is the set of nodes with degree 1 and $\bar\Lambda_g^{(2)}$ is that of nodes with degree 2. It is easy to verify that the numbers of nodes in $\bar\Lambda_g^{(1)}$ and $\bar\Lambda_g^{(2)}$ are
\begin{eqnarray}\label{D00}
|\bar\Lambda_g^{(1)}|=(f-2)(f+1)^{g-1}+2
\end{eqnarray}
and
\begin{eqnarray}\label{D01}
|\bar\Lambda_g^{(2)}|=2(f+1)^{g-1}-2\,,
\end{eqnarray}
respectively.

We now study a new quantity $D_g$, which represents the distance to a given peripheral node in $V_{f,g}$ and is defined by
\begin{eqnarray}\label{C00}
D_g=\sum_{i \in V_{f,g}}d_i(g)\,,
\end{eqnarray}
where $d_{i}(g)$ is the length of the unique shortest path from node $i$ to the peripheral node in $V_{f,g}$.
According to the self-similar structure of $V_{f,g}$, we have the following recursive relation:
\begin{eqnarray}\label{C01}
D_g&=&D_{g-1}+(D_{g-1}+N_{g-1}+N_{g-1}L_{g-1})\nonumber\\
&\quad&+(f-1)(D_{g-1}+2N_{g-1}+2N_{g-1}L_{g-1}).
\end{eqnarray}
Considering $D_1=2f-1$, Eq.~(\ref{C01}) is solved to yield
\begin{eqnarray}\label{C02}
D_g=\frac{1}{2}(3^g-1)(2f-1)(f+1)^{g-1}.
\end{eqnarray}

After introducing the two polymer networks, in what follows we will study the trapping problem defined on them, with an aim to gain a comprehensive understanding of the effect of trap's location on the absorbing efficiency on $C_{m,g}$ and $V_{f,g}$. For this purpose, we will first investigate random walks with a single immobile trap fixed on a certain node; then we will continue to address random walks with the trap distributed uniformly over all nodes.

\section{Trapping with an immobile trap}\label{TrapC}

In this section, we study isotropic random walks with a single trap defined on $C_{m,g}$ and $V_{f,g}$, respectively. The random-walk model considering here is a simple one. At each discrete time step, the walker moves from its current location to an arbitrary nearest neighbors with the same probability. Let $T_{ij}(g)$ denote the MFPT from node $i$ to $j$, which is the expected time taken by a walker starting from $i$ to first reach $j$. Let $T_j(g)$ denote the ATT to trap node $j$. Then, the interesting quantity related to trapping problem is given by
\begin{equation}\label{A00}
T_j(g) = \frac{1}{N_g}\sum_i T_{ij}(g).
\end{equation}
In the sequel, we will study analytically $T_j(g)$ for both $C_{m,g}$ and  $V_{f,g}$, and show how $T_j(g)$ varies with network size.

\subsection{Trapping in Cayley trees with a trap at an arbitrary node}\label{TrapC1}

In a recent paper~\cite{WuLiZhCh12}, we have studied the trapping process on Cayley trees with an immobile trap located on the central node. Here, we attack a general trapping problem on Cayley trees with the trap fixed on an arbitrary node. To attain this goal, we first classify the nodes in $C_{m,g}$ into $g+1$ levels according to their shortest distance to the central node: The central node is at level 0, the nodes created at generation 1 are at level 1, and so on. For convenience of the following description, let $r_i$ represent a node at level $i$, where the subscript $i$ denotes simultaneously the generation it is created at, and its shortest distance to the central node. Note that all nodes at the same level are equivalent to each other, in the sense that the ATT is the same, if any of them is considered as a trap.

Let $T_{r_j}^{\rm sum}(g)$ be the sum of the MFPT from a starting point to a target node at level $j$ ($0\leq j \leq g$) in $C_{m,g}$, where the sum is taken over all starting nodes in $C_{m,g}$. That is,
\begin{eqnarray}\label{A01}
T_{r_j}^{\rm sum}(g)=\sum_{i \in C_{m,g}} T_{i,r_j}(g)\,.
\end{eqnarray}
Then, the ATT to an arbitrary node at level $j$ in $C_{m,g}$ is
\begin{eqnarray}\label{A02}
T_{r_j}(g)=\frac{1}{N_g}T_{r_j}^{\rm sum}(g).
\end{eqnarray}
Thus, to find $T_{r_j}(g)$, we will alternatively evaluate the quantity $T_{r_j}^{\rm sum}(g)$.

In order to obtain $T_{r_j}^{\rm sum}(g)$, we regard $C_{m,g}$ as a rooted tree with the central node being the root. Then, the following relation holds:
\begin{eqnarray}\label{A03}
T_{r_j}^{\rm sum}(g)&=&\left[\frac{m-1}{m}T_{\rm C}^{\rm sum}(g)+\frac{(m-1)^{g+1}-1}{m-2}T_{r_0 r_j}(g)\right]\nonumber\\
&\quad&+\sum_{i=1}^{j-1}\left[\frac{m-2}{m}T_{\rm C}^{\rm sum}(g-i)+(m-1)^{g-i}T_{r_i r_j}(g)\right]\nonumber\\
&\quad&+\frac{m-1}{m}T_{\rm C}^{\rm sum}(g-j),
\end{eqnarray}
where $T_{\rm C}^{\rm sum}(g)$ is the sum of MFPTs for all nodes to the central node on $C_{m,g}$, and $T_{r_i r_j}$ is the MFPT from a node at level $i$ to one of its offspring node at level $j$.

Equation~(\ref{A03}) can be accounted for as follows. The first term is based on the fact that a walker starting from a node, which and the trap node have the only one lowest common ancestor (i.e., the central node), should first visit to the central node, and then takes $T_{r_0 r_j}(g)$ more time steps to reach the trap. Here, the lowest common ancestor for two nodes $i$ and $j$ is defined as the node with the possible biggest level value in the rooted tree but having both $i$ and $j$ as its descendants. The second term describes the case that a particle starting off from a node, which and the trap have the lowest common ancestor at level $i$, first jumps to the lowest common ancestor and then takes $T_{r_i r_j}(g)$ more steps to arrive at the target for the first time. The last term accounts for the sum of MFPTs from all descendants of the trap to the trap itself.

Next we derive the two quantities $T_{\rm C}^{\rm sum}(g)$ and $T_{r_i r_j}(g)$ with $i < j$. In a previous work~\cite{WuLiZhCh12}, we have derived that the MFPT, $T_{r_i {\rm C}}(g)$, from a node at level $i$ to the central node is
\begin{eqnarray}\label{A04}
T_{r_i {\rm C}}(g)=\frac{2}{(m-2)^2}\left[(m-1)^{g+1}-(m-1)^{g-i+1}\right]-\frac{m}{m-2}i,\nonumber\\
\end{eqnarray}
utilizing which we can easily determine $T_{\rm C}^{\rm sum}(g)$ given by
\begin{eqnarray}\label{A05}
T_{\rm C}^{\rm sum}(g)&=&\sum_{i=1}^g N_i(g)T_{r_i {\rm C}}(g)=\frac{m}{(m-2)^3}[2(m-1)^{2g+1}-m]\nonumber\\
&\quad&-\frac{m(m+2)g+m}{(m-2)^2}(m-1)^g.
\end{eqnarray}

We proceed to evaluate $T_{r_i r_j}(g)$ ($i<j$), which can be expressed as
\begin{eqnarray}\label{A06}
T_{r_i r_j}(g)=\sum_{k=i}^{j-1}T_{r_{k} r_{k+1}}(g).
\end{eqnarray}
Before determining $T_{r_i r_j}(g)$, we first calculate the MFPT, $T_{r_i r_{i+1}}(g)$, from a node at level $i$ ($0\leq i < g$) to its neighboring node at level $i+1$.
For the case of $i=0$, it is easy to have
\begin{eqnarray}\label{A07}
T_{r_0 r_1}(g)=\frac{1}{m}+\frac{m-1}{m}\left[1+T_{r_1 \rm C}(g)+T_{r_0 r_1}(g)\right],
\end{eqnarray}
Substituting Eq.~(\ref{A04}) into Eq.~(\ref{A07}), we can solve Eq.~(\ref{A07}) to yield
\begin{eqnarray}\label{A08}
T_{r_0 r_1}(g)=\frac{1}{m-2}\left[2(m-1)^{g+1}-m\right].
\end{eqnarray}
For $i \geq 1$, the following relation holds:
\begin{eqnarray}\label{A09}
T_{r_i r_{i+1}}(g)&=&\frac{1}{m} + \frac{1}{m}\left[1+T_{r_{i-1} r_{i}}(g)+T_{r_{i} r_{i+1}}(g)\right]\nonumber\\
&\quad&+\frac{m-2}{m}\left[1+T_{r_1\rm C}(g-i)+T_{r_{i} r_{i+1}}(g)\right].\nonumber\\
\end{eqnarray}
Considering the initial condition given by Eq.~(\ref{A08}), Eq.~(\ref{A09}) can be solved inductively:
\begin{eqnarray}\label{A10}
T_{r_i r_{i+1}}(g)=\frac{1}{m-2}\left[2m(m-1)^g-2(m-1)^{g-i}-m\right].\nonumber\\
\end{eqnarray}
Inserting Eq.~(\ref{A10}) into Eq.~(\ref{A06}) yields
\begin{eqnarray}\label{A11}
T_{r_i r_j}(g)&=&\sum_{k=i}^{j-1}T_{r_k r_{k+1}}(g)=\frac{m(j-i)}{m-2}[2(m-1)^g-1]\nonumber\\
&\quad&-\frac{2}{(m-2)^2}(m-1)^{g-j+1}[(m-1)^{j-i}-1].\nonumber\\
\end{eqnarray}

Plugging Eq.~(\ref{A05}) and Eq.~(\ref{A11}) into Eq.~(\ref{A03}), we have
\begin{eqnarray}\label{A12}
T_{r_j}^{\rm sum}(g)&=&\frac{2m(m-1)^{2g}}{(m-2)^3}[jm(m-2)-m+1+2(m-1)^{1-j}]\nonumber\\
&\quad&+\frac{(m-1)^g}{(m-2)^3}[(g+j)(4m-m^3)+m^2+4m-4]\nonumber\\
&\quad&-\frac{2(m+2)(m-1)^{g-j+1}}{(m-2)^3}-\frac{m^2}{(m-2)^3}.
\end{eqnarray}
Substituting the expression of Eq.~(\ref{A12}) into Eq.~(\ref{A02}), we obtain the rigorous expression for the ATT to the trap node at level $j$ on the $g$th generation of the Cayley trees:
\begin{eqnarray}\label{A13}
 T_{r_j}(g)&=&\frac{2m(m-1)^{2g}[jm(m-2)-m+1+2(m-1)^{1-j}]}{(m-2)^2[m(m-2)^g-2]}\nonumber\\
&\quad&+\frac{(m-1)^g[(g+j)(4m-m^3)+m^2+4m-4]}{(m-2)^2[m(m-2)^g-2]}\nonumber\\
&\quad&-\frac{2(m+2)(m-1)^{g-j+1}+m^2}{(m-2)^2[m(m-2)^g-2]}.
\end{eqnarray}
Equation~(\ref{A13}) provides an explicit formula for ATT to an arbitrary node of $C_{m,g}$. For the particular case of $j=0$ that the trap is fixed on the central node, Eq.~(\ref{A13}) is reduced to the previous result~\cite{WuLiZhCh12}.
For another limiting case ($j=g$) that a specific peripheral node is looked upon the trap, by substituting $j=g$ into Eq.~(\ref{A13}), we obtain the ATT as
\begin{eqnarray}\label{A15}
 T_{r_g}(g)&=&\frac{2m^2(m-2)g-2m(m-1)}{(m-2)^2[m(m-1)^g-2]}(m-1)^{2g}\nonumber\\
&\quad&-\frac{2m(m^2-4)g-5m^2+4}{(m-2)^2[m(m-1)^g-2]}(m-1)^g\nonumber\\
&\quad&-\frac{3m^2+2m-4}{(m-2)^2[m(m-1)^g-2]}.
\end{eqnarray}

We continue to express $T_{r_j}(g)$ in terms of the network size $N_g$, in order to obtain the relation governing the two quantities. Recalling Eq.~(\ref{NodeC}), we have $(m-1)^g=[(m-2)N_g+2]/m$ and $g=[\ln((m-2)N_g+2)-\ln m]/\ln(m-1)$. These relations allow to recast $ T_{r_j}(g)$ as a function of $N_g$:
\begin{eqnarray}\label{A16}
T_{r_j}(g)&=&\frac{2[jm(m-2)-m+1]}{m(m-2)^3N_g}[(m-2)N_g+2]^2\nonumber\\
&\quad&+\frac{4}{m(m-1)^{j-1}(m-2)^3N_g}[(m-2)N_g+2]^2\nonumber\\
&\quad&+\frac{(m^2-4)[1-(g+j)m]+4m}{m(m-2)^3N_g}[(m-2)N_g+2]\nonumber\\
&\quad&-\frac{2(m+2)[(m-2)N_g+2]}{m(m-1)^{j-1}(m-2)^3N_g}-\frac{m^2}{(m-2)^3N_g}.
\end{eqnarray}
Thus, for a very large system, i.e., $N_g\rightarrow \infty$, we have the following expression
for the leading term of $T_{r_j}(g)$:
\begin{equation}\label{A17}
T_{r_j}(g)\sim (j+1)N_g\,.
\end{equation}

Equation~(\ref{A17}) shows that leading behavior of $T_{r_j}(g)$ is dependent on the level of trap position, i.e., the distance from the central node to the trap.
Particularly, when $j=0$, namely the central node is trap, we have $T_{r_j}(g)\sim N_g$; for $j=g$ that trap is located at a peripheral node, Eq.~(\ref{A17}) implies $T_{r_j}(g)\sim N_g\ln N_g$. Thus, in the limit of the large network size $N_g$, for the trap located on the central node, the ATT grows linearly with the increasing network size. However, for the trap fixed on a particular peripheral node, the leading asymptotic $N_g\ln N_g$ dependence of $ T_{r_g}(g)$ with the network size is in strong contrast with the linear scaling of $T_{r_0}(g)$ with $N_g$.

\subsection{Trapping in Vicsek fractals with the trap at a peripheral node}

Different from the case of Cayley trees, for trapping problem in Vicsek fractals $V_{f,g}$, it is very difficult and even impossible to determine the exact expression for ATT when the trap is located at an arbitrary node. But for some cases that the trap is fixed at a particular node, the problem can be solved analytically. In a previous work~\cite{WuLiZhCh12}, we have obtained the ATT to the central node on $V_{f,g}$. Below we will utilize a similar but a little different technique to address the trapping process on $V_{f,g}$ with an immobile trap positioned on a specific peripheral node. We will show that, for both cases, the leading behavior for ATT is identical.

Let $T_{i\rm P}(g)$ denote the TT for node $i$, i.e., the MFPT from node $i$ to the trap node. Then, for this case the ATT $ T_{\rm P}(g)$ is given by
\begin{eqnarray}\label{I02}
T_{\rm P}(g)=\frac{1}{N_g}\sum_{i \in \Lambda_g}{T_{i\rm P}(g)}\,.
\end{eqnarray}
In order to determine $T_{i\rm P}(g)$, we introduce two new quantities for $n\leq g$:
\begin{eqnarray}\label{C05}
T_{n}^{\rm sum}(g)=\sum_{i\in \Lambda_n}T_{i\rm P}(g)
\end{eqnarray}
and
\begin{eqnarray}\label{C06}
{\bar T}_n^{\rm sum}(g)=\sum_{i\in\bar\Lambda_n}T_{i\rm P}(g).
\end{eqnarray}
Then, we have
\begin{eqnarray}\label{I022}
T_{\rm P}(g)=\frac{1}{N_g}T_{g}^{\rm sum}(g)\,
\end{eqnarray}
and
\begin{eqnarray}\label{C07}
T_g^{\rm sum}(g)=T_{g-1}^{\rm sum}(g)+{\bar T}_g^{\rm sum}(g).
\end{eqnarray}
In this way, the problem of determining $ T_{\rm P}(g)$ is reduced to finding $T_{g-1}^{\rm sum}(g)$ and ${\bar T}_g^{\rm sum}(g)$.

We first deduce the reclusive relation for $T_{g-1}^{\rm sum}(g)$. Using a similar process as the case that the trap is fixed on the central node~\cite{WuLiZhCh12}, we can derive the following law governing the evolution for TT of node $i$:
\begin{eqnarray}\label{B03}
T_{i \rm P}(g)=3(f+1)T_{i\rm P}(g-1)+3f d_i(g-1)\,.
\end{eqnarray}
Note that according to the second construction approach, for a particle performing random walks in $V_{f,g}$, before arriving at the peripheral node as the trap, it must first visit its unique neighbor that is in fact a peripheral node on $V_{f,g-1}$. Then, we have
\begin{eqnarray}\label{C08}
T_{g-1}^{\rm sum}(g)&=&\sum_{i\in\Lambda_{g-1}}T_{i\rm P}(g)\nonumber\\
&=&\sum_{i\in\Lambda_{g-1}}[3(f+1)T_{i\rm P}(g-1)+3f d_i(g-1)]\nonumber\\
&\quad&+N_{g-1}[2(N_g-1)-1]\,.
\end{eqnarray}
The sum term on the right-hand side (rhs) of Eq.~(\ref{C08}) stands for the time spent by a walker to reach the neighbor of the trap, while the second term accounts for the time steps from the trap's neighbor to the trap. Equation~(\ref{C08}) can be readily simplified to
\begin{eqnarray}\label{C09}
T_{g-1}^{\rm sum}(g)&=&3(f+1)T_{g-1}^{\rm sum}(g-1)+3f D_{g-1}+N_{g-1}(2N_g-3)\,.\nonumber\\
\end{eqnarray}

For ${\bar T}_g^{\rm sum}(g)$, it can be evaluated as follows.
By definition,
\begin{eqnarray}\label{C10}
{\bar T}_g^{\rm sum}(g)=\sum_{i\in\bar\Lambda_g^{(1)}}T_{i\rm P}(g)+\sum_{i\in\bar\Lambda_g^{(2)}}T_{i\rm P}(g)\,.
\end{eqnarray}
Applying the approach in~\cite{WuLiZhCh12}, we can evaluate the two summation terms on the rhs of Eq.~(\ref{C10}), and further obtain the following recursive relation for ${\bar T}_g^{\rm sum}(g)$:
\begin{eqnarray}\label{CC17}
{\bar T}_g^{\rm sum}(g)=f T_{g-1}^{\rm sum}(g)+4E_{g-1}+|\bar\Lambda_g^{(1)}|-2N_g+2\,.
\end{eqnarray}

Plugging Eqs.~(\ref{C09}) and Eq~(\ref{CC17}) into Eq.~(\ref{C07}) gives
\begin{eqnarray}\label{C18}
T_g^{\rm sum}(g)&=&3(f+1)^2T_{g-1}^{\rm sum}(g-1)+3f(f+1)D_{g-1}\nonumber\\
&\quad&+N_{g-1}(f+1)(2N_g-3)+|\bar\Lambda_g^{(1)}|\nonumber\\
&\quad&+4E_{g-1}-2N_g+2\,.
\end{eqnarray}
Considering $T_1^{\rm sum}(1)=2f^2-1$ and combining the above-obtained results, Eq.~(\ref{C18}) can be solved to obtain
\begin{eqnarray}\label{C19}
T_g^{\rm sum}(g)&=&\frac{3f^2+3f-2}{3f^2+5f+2}3^g(f+1)^{2g}-(f+1)^{2g}\nonumber\\
&\quad&-\frac{2f-1}{2f+2}3^g(f+1)^g+\frac{6f^2+5f+6}{6f^2+10f+4}(f+1)^g.\nonumber\\
\end{eqnarray}

Inserting Eq.~(\ref{C19}) into Eq.~(\ref{I022}), we arrive at the exact formula for the ATT with the trap located at a specified peripheral node on $V_{f,g}$:
\begin{eqnarray}\label{C20}
 T_{\rm P}(g)&=&\frac{3f^2+3f-2}{3f^2+5f+2}3^g(f+1)^{g}-(f+1)^{g}\nonumber\\
&\quad&-\frac{2f-1}{2f+2}3^g+\frac{6f^2+5f+6}{6f^2+10f+4}\,,
\end{eqnarray}
which can be further represented as a function of network size $N_g$ as
\begin{eqnarray}\label{C21}
T_{\rm P}(g)&=&\frac{3f^2+3f-2}{3f^2+5f+2}(N_g)^{1+\log_{3}(f+1)}-N_g\nonumber\\
&\quad&-\frac{2f-1}{2f+2}(N_g)^{\log_{3}(f+1)}+\frac{6f^2+5f+6}{6f^2+10f+4}\,.
\end{eqnarray}
From this succinct dependence relation of $T_{\rm P}(g)$ on network size $N_g$, we can find that for large networks, i.e., $N_g\to\infty$, the leading term is
\begin{eqnarray}\label{C22}
T_{\rm P}(g)\sim (N_g)^{1+\log_{3}(f+1)}\,,
\end{eqnarray}
which is identical to the behavior of ATT $T_{\rm C}(g)$ when the trap is located at the central node~\cite{WuLiZhCh12}.

\section{Trapping with the trap uniformly distributed}

In Sec.~\ref{TrapC}, we have studied the trapping problem on $C_{m,g}$ and $V_{f,g}$, with an immobile trap located at a given node. In this section, we will study the trapping issue on the two networks with the trap uniformly distributed throughout all nodes.

In this case, what we are concerned with is the quantity $T_g$ defined as the average of MFPTs over all pairs of nodes in the networks:
\begin{eqnarray}\label{E00}
T_g=\frac{1}{(N_g)^2}\sum_{i=1}^{N_g}\sum_{j=1}^{N_g} T_{ij}(g)\,.
\end{eqnarray}
For convenience, we use $T_{\rm tot}(g)$ to denote the summation term on the rhs of  Eq.~(\ref{E00}):
\begin{eqnarray}\label{E01}
T_{\rm tot}(g)=\sum_{i=1}^{N_g}\sum_{j=1}^{N_g} T_{ij}(g)\,.
\end{eqnarray}
Then,
\begin{eqnarray}\label{E000}
T_g=\frac{T_{\rm tot}(g)}{(N_g)^2}\,,
\end{eqnarray}
which is called global average tapping time (GATT).

By definition, the quantity GATT involves a double average: The first one is over all the starting nodes to a given trap node, the second one is the average of the first one with the trap having a uniform distribution among all nodes. In the sequel, we will analytically study $T_g$ for Cayley trees $C_{m,g}$ and Vicsek fractals $V_{f,g}$, respectively. For convenience, hereafter we also call $T_g$ the ATT in the case without confusion.

\subsection{Cayley trees}

For Cayley trees, we can easily determine $T_{\rm tot}(g)$ and $T_g$ by using the intermediary results obtained in Section~\ref{TrapC1}. For $T_{\rm tot}(g)$, it obeys the following relation:
\begin{eqnarray}\label{F01}
T_{\rm tot}(g)=T_{\rm C}^{\rm sum}(g) + \sum_{j=1}^{g}N_j(g)T_{r_j}^{\rm sum}(g)\,.
\end{eqnarray}
Substituting Eq.~(\ref{A12}) into Eq.~(\ref{F01}), we obtain
\begin{eqnarray}\label{F05}
T_{\rm tot}(g)&=&\frac{2m^2}{(m-2)^4}[(m-1)^g-1][2m(m-1)^g-1\nonumber\\
&\quad&+((m^2-2m)g-2m+1)(m-1)^{2g}]\,.
\end{eqnarray}
Then, the explicit expression for $T_g$ in $C_{m,g}$ is
\begin{eqnarray}\label{F06}
T_g&=&\frac{2m^2}{(m-2)^2[m(m-1)^g-2]^2}[(m-1)^g-1][2m(m-1)^g\nonumber\\
&\quad&+((m^2-2m)g-2m+1)(m-1)^{2g}-1]\,,\nonumber\\
\end{eqnarray}
which can be rewritten in terms of the network size $N_g$ as
\begin{widetext}
\begin{eqnarray}\label{F07}
 T_g&=&\frac{2(N_g-1)[(m-2)N_g+2]^2}{m(m-2)^3(N_g)^2}\left[\frac{\ln(m N_g-2N_g+2)-\ln m}{\ln(m-1)}(m^2-2m)-2m+1\right]+\frac{4m(N_g-1)}{(m-2)^2N_g}+\frac{6m(N_g-1)}{(m-2)^3(N_g)^2}.\nonumber\\
\end{eqnarray}
\end{widetext}
In the limit of infinite network size (i.e., $N_g \to \infty$), we have the dominating term
\begin{eqnarray}\label{F08}
T_g \sim N_g\ln N_g.
\end{eqnarray}
This leading asymptotic $N_g \ln N_g$ dependence of $T_g$ on the network size is equivalent to that of $T_{r_g}(g)$ for trapping in $C_{m,g}$ with a peripheral node being a trap, but is in marked contrast with the
linear scaling of $T_{r_0}(g)$ for the trapping problem when the central node is the trap.

\subsection{Vicsek fractals}

For Vicsek fractals, the above method for computing
$T_g$ in $C_{m,g}$ is not applicable to
that in $V_{f,g}$. We next resort to another method for determining $T_g$ in $V_{f,g}$, by using the connection~\cite{ChRaRuSm89,Te91} between resistance distance, also refereed to as effective resistance, and MFPTs for random walks on a connected graph. To this end, we view $V_{f,g}$ as an electrical network~\cite{DoSn84} by considering each edge in $V_{f,g}$ to be a unit resistor~\cite{KlRa93}. Let $R_{ij}(g)$ be the effective resistance between two nodes $i$ and $j$ in the electrical network corresponding to $V_{f,g}$. Then, we have following exact relation~\cite{ChRaRuSm89,Te91}
\begin{eqnarray}\label{H00}
T_{ij}(g)+T_{ji}(g)=2E_g\, R_{ij}(g)\,.
\end{eqnarray}
Using this obtained relation governing MFPTs and effective resistance, Eq.~(\ref{E01}) can be recast as
\begin{eqnarray}\label{H01}
T_{\rm tot}(g)=E_g\sum_{i=1}^{N_g}\sum_{j=1}^{N_g}R_{ij}(g)\,.
\end{eqnarray}

Equation~(\ref{H01}) tells us that if we have a method to determine the effective resistance, then we can find the quantity $T_g$. Since Vicsek fractals have a treelike structure, the effective resistance $R_{ij}(g)$ is exactly the usual shortest-path distance between node $i$ and $j$ in $V_{f,g}$, which we denote as $s_{ij}(g)$. Then,
\begin{eqnarray}\label{H02}
F_{\rm tot}(g)=E_g\sum_{i=1}^{N_g}\sum_{j=1}^{N_g}s_{ij}(g)=E_g\,S_{\rm tot}(g),
\end{eqnarray}
where
\begin{eqnarray}\label{H03}
S_{\rm tot}(g)=\sum_{i=1}^{N_g}\sum_{j=1}^{N_g}s_{ij}(g)\,
\end{eqnarray}
is actually the Winner index~\cite{Wi47,DoenGu01} of $V_{f,g}$.

We continue by showing the procedure of determining the total shortest-path distance $S_{\rm tot}(g)$, which just equals the number of edges in the shortest paths between all pairs of nodes in $V_{f,g}$. Instead of counting the edges in the paths, here we count the paths passing through a given edge, and then sum the results of all edges in $V_{f,g}$. Let $(i,j)$ be an edge in $V_{f,g}$ connecting nodes $i$ and $j$, and $E_{ij}(g)$ the number of the shortest paths of different node pairs, which pass through $(i,j)$. Let $N_{i<j}(g)$ be the number of nodes in $V_{f,g}$ lying closer to node $i$ than to node $j$, including $i$ itself. Then,
\begin{eqnarray}\label{H04}
S_{\rm tot}(g)&=&\sum_{(i,j)\in V_{f,g}}E_{ij}(g)=\sum_{(i,j)\in V_{f,g}}2N_{i<j}(g)N_{j<i}(g)\nonumber\\
&=&2\sum_{(i,j)\in V_{f,g}}N_{i<j}(g)[N_g-N_{i<j}(g)]\,,
\end{eqnarray}
where $N_{j<i}(g)=N_g-N_{i<j}(g)$ was made use of.

We now apply the relation in Eq.~(\ref{H04}) to deduce $S_{\rm tot}(g)$.
For this purpose, we classify the edges in $V_{f,g}$ into two sets in the following way. Recalling the second construction of $V_{f,g}$ (see Fig.~\ref{Const2}), an arbitrary edge $(u,v)$ in $V_{f,g-1}$ is replaced by three new edges $(u, u_1)$, $(u_1, v_1)$ and $(v_1, v)$. Let $e_g^{(1)}$ denote the set of those edges in $V_{f,g}$, with both endpoints of each edge connecting two nodes having a degree more than one. And let $e_g^{(2)}$ denote set of the remaining edges, all of which have exactly an endpoint with a single degree. It is easy to derive that the numbers of edges in these two sets are
\begin{eqnarray}\label{H05}
|e_g^{(1)}|=3E_{g-1}=3(f+1)^{g-1}-3
\end{eqnarray}
and
\begin{eqnarray}\label{H06}
|e_g^{(2)}|=E_g-|e_g^{(1)}|=(f-2)(f+1)^{g-1}+2\,,
\end{eqnarray}
respectively.

Figure~\ref{Const2} implies that for an edge $(u,v)$ in $V_{f,g-1}$, we have the following relation for the three edges $(u,u_1)$, $(u_1,v_1)$, and $(v_1,v)$, in $V_{f,g}$:
\begin{eqnarray}\label{H07}
E_{uu_1}(g)+E_{u_1v_1}(g)+E_{v_1v}(g)=3(f+1)^2E_{uv}(g-1)-4\,.\nonumber\\
\end{eqnarray}
Summarizing the terms on the rhs of Eq.~(\ref{H07}) over all the $E_{g-1}$ edges in $V_{f,g-1}$ yields
\begin{eqnarray}\label{H08}
\sum_{(i,j)\in e_{g}^{(1)}}E_{ij}(g)=3(f+1)^2 S_{\rm tot}(g-1)-4E_{g-1}.
\end{eqnarray}
On the other hand, for edges in $e_{g}^{(2)}$, we have
\begin{eqnarray}\label{H09}
\sum_{(i,j)\in e_{g}^{(2)}}E_{ij}(g)=2|e_{g}^{(2)}|(N_g-1).
\end{eqnarray}

Making use of Eqs.~(\ref{H08}) and~(\ref{H09}), the quantity $S_{\rm tot}(g)$ can be represented recursively as
\begin{eqnarray}\label{H10}
S_{\rm tot}(g)&=&\sum_{(i,j)\in e_g^{(1)}}E_{ij}(g)+\sum_{(i,j)\in e_g^{(2)}}E_{ij}(g)\nonumber\\
&=&3(f+1)^2S_{\rm tot}(g-1)-4E_{g-1}+2|e_{g}^{(2)}|(N_g-1).\nonumber\\
\end{eqnarray}
Considering $S_{\rm tot}(1)=2f^2$, Eq.~(\ref{H10}) can be solved inductively to obtain
\begin{eqnarray}\label{H14}
S_{\rm tot}(g)&=&\frac{(f+1)^{g-1}}{3f+2}[(f+1)^g((3f^2-2 f)3^g-3f^2+4f+4)\nonumber\\
&\quad&-2(f+2)].
\end{eqnarray}
Then, the analytical expression for $T_g$ in $V_{f,g}$ is
\begin{eqnarray}\label{H15}
T_g&=&\frac{E_g S_{\rm tot}(g)}{(N_g)^2}=\frac{[(f+1)^g-1]}{(3f+2)(f+1)^{g+1}}[-2(f+2)\nonumber\\
&\quad&+(f+1)^g((3f^2-2 f)3^g-3f^2+4f+4)]\,,
\end{eqnarray}
which can be rewritten explicitly in terms of  network size $N_g$ as
\begin{eqnarray}\label{H16}
T_g&=&\frac{N_g-1}{(3f+2)N_g}[(3f^2-2 f)(N_g)^{1+\log_{3}(f+1)}\nonumber\\
&\quad&-N_g(3f^2+4f+4)-2(f+2)]\,.
\end{eqnarray}
When $N_g\to\infty$, we have the dominating term for $T_g$ in $V_{f,g}$:
\begin{eqnarray}\label{H17}
T_g\sim (N_g)^{1+\log_{3}(f+1)}\,,
\end{eqnarray}
growing as a power-law function of network size $N_g$, a behavior similar to that of $T_{\rm P}(g)$.

Note that the above method and process for computing Winner index and ATT is general for all trees. We have used this approach to calculate $T_g$ for Cayley trees and recovered the result in Eq.~(\ref{F06}).

\section{Result comparison and analysis}

From the result provided by Eq.~(\ref{A17}), we can easily see that the leading behaviors of $ T_{r_j}(g)$ for different $j$ are evidently different. This distinction shows that for trapping in Cayley trees $C_{m,g}$ with a deep trap, the trap's location has a significant effect on the trapping efficiency measured by ATT. Specifically, the dominating scaling of ATT grows with the distance from trap to the central node: the smaller the distance, the more efficient the trapping process. For the case that the trap is the central node, the trapping process is the most efficient, with the scaling of ATT growing linearly with system size $N_g$; while for the case when the trap is placed at a peripheral node, the trapping process is the least efficient, with the ATT changing with network size $N_g$ as $N_g\ln N_g$.

In addition, Eq.~(\ref{F08}) shows that when the trap is uniformly distributed over the whole Cayley trees, the dominating scaling of $T_g$ behaves with the network size $N_g$ as $N_g\ln N_g$. This indicates that the linear scaling of the ATT $T_{r_0}(g)$ to the central node is not representative of the Cayley trees, in the sense that $T_g$ scales larger than linearly. In contrast, the scaling of ATT to a peripheral node is a representative property for trapping process taking place in Cayley trees.

The different scalings for $T_{r_j}(g)$ lie in the peculiar structure of Cayley trees. Figure~\ref{Cayley} shows that a Cayley tree actually consists of $m$ branches (regions), each of which is a subtree with a node at level $1$ being its root. For $j=0$ corresponding to the case when the trap node is the central node, the walker, irrespective of its starting point, will visit at most one branch before being trapped. On the contrary, for $j>0$ corresponding the case that the trap is located on a node with distance $j$ to the central node, the particle, starting from a large group of nodes, must first visit the central node and then proceeds from the central node along the path $r_{0}- r_{1}-\cdots -r_{j-1}-r_{j}$ until it is absorbed by the trap.
According to previous result~\cite{LiJuZh12}, the MFPT from a node at level $j-1$ to its direct neighbor at level $j$ is $T_{r_{j-1}r_j}=2N_{r_{j-1}<r_j}-1$, which increases with $j$. Thus, the scaling of $T_{r_j}(g)$ grows with $j$, as shown in Eq.~(\ref{A17}). For instance, for the two limiting cases of $j=0$ and $j=g$, $T_{r_0}(g) \sim N_g $ and $ T_{r_g}(g) \sim N_g\ln N_g $ since $g \sim \ln N_g$. Finally, for the case that the trap is uniformly distributed, the phenomenon that the leading term of $T_g$ exhibits the same scaling as that of $T_{r_g}(g)$ can be heuristically understood as follows. From Eqs.~(\ref{PeriC}) and (\ref {NodeC}), it is easy to see that in $C_{m,g}$ the fraction of peripheral nodes is about $\frac{m-2}{m-1}$, thus $T_{r_g}(g)$ alone can determine the behavior of $T_g$.

\begin{figure}
\begin{center}
\includegraphics[width=1.15\linewidth,trim=20 30 0 30]{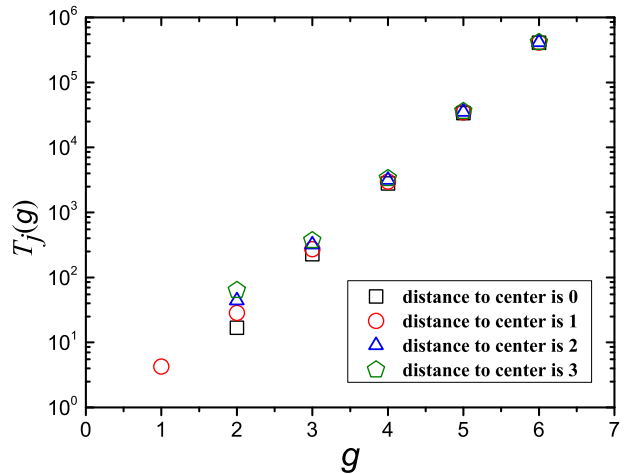}
\end{center}
\caption[kurzform]{(Color online) Average trapping time versus the distance from the trap to central node in Vicsek fractals corresponding to $f=3$.}\label{ATT}
\end{figure}

Quite different from those in Cayley trees, the dominating scaling of $T_{\rm C}(g)$, $T_{\rm P}(g)$, and $T_g$ in Vicsek fractals are identical, all of which scale with $N_g$ as $(N_g)^{1+\log_{3}(f+1)}$ as shown in Eqs.~(\ref{C22}) and~(\ref{H17}). Moreover, extensive numerical results also show that for case that the trap is placed at other node, the leading behavior for ATT is also $N_g$ as $(N_g)^{1+\log_{3}(f+1)}$. In Fig.~\ref{ATT}, we report the numerical results for trapping in $V_{3,g}$ ($1 \leq g \leq 7$) with the trap at different nodes, which shows that the ATTs display the same scaling, independent of the trap's location. Therefore, the location of the trap has no qualitative effect on the scaling of ATT for trapping in Vicsek fractals, which is in marked contrast to the trapping process occurring on Cayley trees.

The root of identical scaling for ATTs corresponding to different positions of the trap is attributed to the structure of Vicsek fractals. For the case that the trap is at the central node~\cite{LiJuZh12}, the ATT is the lowest, the leading scaling of which varies with network size $N_g$ as $N_g^{1+\log_{3}(f+1)}$. For the case that the trap is at another node different from the central node, the analysis is similar to that of Cayley trees. In this case, to find the target in $V_{f,g}$, the walker must first visit the central node, taking $(N_g)^{1+\log_{3}(f+1)}$ time steps, then continues to jump towards the trap along the unique shortest path from the central node to the trap. During the process of jumping along the path, the MFPT from a node $x$ to its direct neighbor $y$ is $2N_{x<y}-1$, which is proportional to $N_g$. Particularly, for the special case when the trap is positioned at a peripheral node, the ATT is the highest, which is approximately equal to $N_g\times L_g/2\sim (N_g)^{1+\log_{3}(f+1)}$, where  $L_g=3^g-1 \sim (N_g)^{\log_{3}(f+1)}$ is the diameter of $V_{f,g}$, as defined above. Since when the trap is fixed at another node other than the central node and peripheral nodes, the ATT is between $ T_{\rm C}(g)$ and $T_{\rm P}(g)$, and thus display the same scaling as $T_{\rm C}(g)$ and $ T_{\rm P}(g)$. Analogously, we can account for the leading asymptotic dependence $(N_g)^{1+\log_{3}(f+1)}$ of $T_g$ corresponding to the case when the trap is uniformly distributed over all nodes.

From the aforementioned results, it can be seen that the trapping efficiency in Cayley trees exhibits rich scalings in the context of the ATT when the trap is placed at different positions. However, the trapping efficiency is identical for Vicsek fractals, despite of the trap's location. In addition to this distinction, there are some other differences between the ATT for these two networks. For example, when the trap is the central node, the trapping efficiency for Cayley trees is higher than that of Vicsek fractals~\cite{WuLiZhCh12}. Moreover, by comparing Eq.~(\ref{A17}) and Eq.~(\ref{C22}), we know that for the case of trap being located at a peripheral node, the ATT of Cayley trees is also much smaller than that of Vicsek fractals. Finally, for the case that the trap is uniformly distributed, the trapping process in Cayley trees is also more efficient in Vicsek fractals, which can be seen by comparing Eq.~(\ref {F08}) with Eq.~(\ref {H17}). Thus, in comparison with Vicsek fractals, Cayley trees have a desirable structure favorable to the trapping process.

It has been proven~\cite{LiJuZh12} that for trapping problem in a general connected graph with an absorbing node, the possible minimal scaling for ATT is proportional to the graph size and the inverse degree of absorbing node, which provides a maximal scaling for the lower bound of ATT for trapping in an arbitrary network with a perfect trap. In this sense, for trapping in Cayley trees, the possible minimal scaling of the ATT can only be reached when the trap is located at the central node, since in this case the leading scaling of the ATT grows proportionally to the system size and the reciprocal of the degree of the central node; while for the case when any other node is the trap, the magnitude of the ATT is greater than the corresponding possible minimal scaling. This phenomenon is in sharp contrast to that observed for the hierarchical scale-free graph~\cite{AgBu09,AgBuMa10,MeAgBeVo12,YaZh13}, where the possible minimal scaling of ATT can be achieved for any node. In the context of Vicsek fractals, the possible minimal scaling cannot be reached for any trapping node, which is also different from that for Cayley trees.

The above result implies that for trapping in a graph, although the possible minimal scaling can be reached for a given trap, it may not be achieved when the trap is placed on another node. In fact, the possible minimal scaling for ATT to a trap depends not only on the  position of the trap, but also on the whole topological structure of the graph~\cite{LiJuZh12}. In future study, it is interesting to explore the problem of designing networks with desirable architecture, for trapping process taking place on which the possible minimal scaling can be achieved for any node as a trap.

\section{Conclusions}

We have performed an in-depth study of the trapping problem on two classical polymer networks---Cayley trees and Vicsek fractals---with a goal to reveal the influence of trap's location on the trapping efficiency. To this aim, for both networks, we first studied the trapping problem with the trap located at a given node; then we addressed the case with the trap uniformly distributed over all nodes in the networks. For both cases of trapping problems, we studied the ATT as an indicator of the trapping efficiency.

We showed that although for a general graph, computing ATT to an arbitrary trap is a theoretical challenge, for Cayley trees, the ATT to any node can be explicitly determined, whose leading term is an increasing function of the shortest distance between the central node and the trap, implying that the place of trap plays an important role in the trapping efficiency. However, for Vicsek fractals, it is very hard and even impossible to obtain the exact expression of ATT for an arbitrary trap, we can only obtain the analytic closed-form formula for ATT to a peripheral node or the central node, while provided numerical results for the ATT to other node, with both analytic and numerical solutions obeying the same leading scaling. Thus, different from the case in Cayley trees, the trap position has little impact on the trapping efficiency for Vicsek fractals.

For the trapping problem when the trap is uniformly distributed, we determined the explicit expressions for ATT in both Cayley trees and Vicsek fractals. The obtained results show that trapping process in Cayley trees is much efficient than in Vicsek fractals. In addition, we also compared the differences of other aspects for trapping in the two networks. Finally, we demonstrated that the root of all differences for trapping in Cayley trees and Vicsek fractals is attributed to their structures.

\begin{acknowledgments}
The authors thank Bin Wu for his assistance in preparing this manuscript. This work was supported by the National Natural Science Foundation
of China under Grant Nos. 61074119 and 11275049.
\end{acknowledgments}

\nocite{*}

\end{document}